\documentclass[twocolumn,prb,aps,10pt]{revtex4-1}

\usepackage{amsfonts,amssymb}
\usepackage[sumlimits,intlimits]{amsmath}
\usepackage{graphics}
\usepackage{graphicx}
\usepackage{mathrsfs}
\usepackage{textcomp}
\usepackage{verbatim}
\usepackage{hyperref}    
\usepackage{bm}          

\pdfoutput=1

\newcommand{\be}{\begin{equation}}
\newcommand{\ee}{\end{equation}}
\newcommand{\lb}{l_B}
\newcommand{\efo}{E_\text{FO}}
\newcommand{\num}{\nu_\text{m}}
\newcommand{\nuc}{\nu_\text{c}}
\newcommand{\ecoh}{E_\text{coh}}

\begin{document}

\title{Giant capacitance of a plane capacitor with a two-dimensional electron gas in a magnetic field}

\author{Brian Skinner}
\author{B. I. Shklovskii}
\affiliation{Fine Theoretical Physics Institute, University of Minnesota, Minneapolis, MN 55455, USA}

\date{\today}

\begin{abstract}

If a clean two-dimensional electron gas (2DEG) with small concentration $n$ comprises one electrode of a plane capacitor, the resulting capacitance $C$ can be larger than the ``geometric capacitance" $C_g$ determined by the physical separation $d$ between electrodes.  A recent paper~\cite{Skinner2010alc} argued that when the effective Bohr radius $a_B$ of the 2DEG satisfies $a_B \ll d$, one can achieve $C \gg C_g$ at low concentration $nd^2 \ll 1$.  Here we show that even for devices with $a_B > d$, including graphene, for which $a_B$ is effectively infinite, one also arrives at $C \gg C_g$ at low electron concentration if there is a strong perpendicular magnetic field.

\end{abstract}

\maketitle

\section{Introduction} \label{sec:intro}

In a standard parallel-plate capacitor, the capacitance per unit area $C$ is equal to the ``geometric capacitance" $C_g = \varepsilon /4 \pi d$ (in Gaussian units), where $\varepsilon$ is the dielectric constant of the medium separating the two plates and $d$ is the distance between them.  The relation $C = C_g$ is correct when both electrodes are made from a perfect metal, which by definition has an infinite electron density of states.  On the other hand, when one of the capacitor electrodes has a finite density of states, there is an additional ``quantum capacitance" contribution to the total capacitance that reflects the finite thermodynamic compressibility of the electronic charge.  This quantum capacitance $C_q$ can be written $C_q = e^2 dn/d\mu$, where $e$ is the electron charge, $n$ is the two-dimensional electron concentration on the electrode surface, and $\mu$ is the electron chemical potential.  Thus, the total capacitance of a capacitor containing one ideal metal electrode and one electrode with finite density of states can be written
\be
C^{-1} = C_g^{-1} + \frac{1}{e^2} \frac{d\mu}{dn}. \label{eq:Cdef}
\ee
More generally, this capacitance is defined as $C = e dn/dV$, where $V$ is the difference in electrochemical potential between the two electrodes, maintained by some external voltage source.
In the following discussion it is convenient to define the effective capacitor thickness $d^* = \varepsilon/4\pi C$, so that when the thermodynamic density of states (TDOS) $d n/d \mu$ is positive, the effective thickness $d^*$ of the capacitor is larger than the physical thickness $d$.

On the other hand, it has long been understood that for a low-density two-dimensional electron gas (2DEG), the TDOS can be \emph{negative} \cite{Bello1981dol, Luryi1988qcd, Kravchenko1990eio, Efros1990dos, Eisenstein1992nco, Eisenstein1994cot, Shapira1996toc, Dultz2000tso, Ilani2000ubo, Allison2006tdo, Shklovskii1986soo, Efros1992hai, Efros1992tdo, Pikus1993doe, Shi2002dsa, Fogler2004nsa, Efros2008ndo, Kopp2009coc}, owing to the emergence of strong positional correlations between electrons.  This implies a negative quantum capacitance, or in other words, $d^* < d$, as was first measured experimentally over two decades ago \cite{Kravchenko1990eio, Eisenstein1992nco, Eisenstein1994cot, Shapira1996toc, Dultz2000tso, Ilani2000ubo}.  Recent experiments on ultra-thin capacitor devices have reported a very large negative quantum capacitance, so that $d^*/d$ is as small as\cite{Li2011vlc, Tinkl2012lne} $0.6$.  These experiments, along with the recent development of nanometer-thick graphene capacitor devices \cite{Ponomarenko2011tmt, Gorbachev2012scd, Sanchez-Yamagishi2012qhe}, bring to the forefront an important and fundamental question: how small can $d^*$ be?

\begin{figure}[htb!]
\centering
\includegraphics[width=0.45 \textwidth]{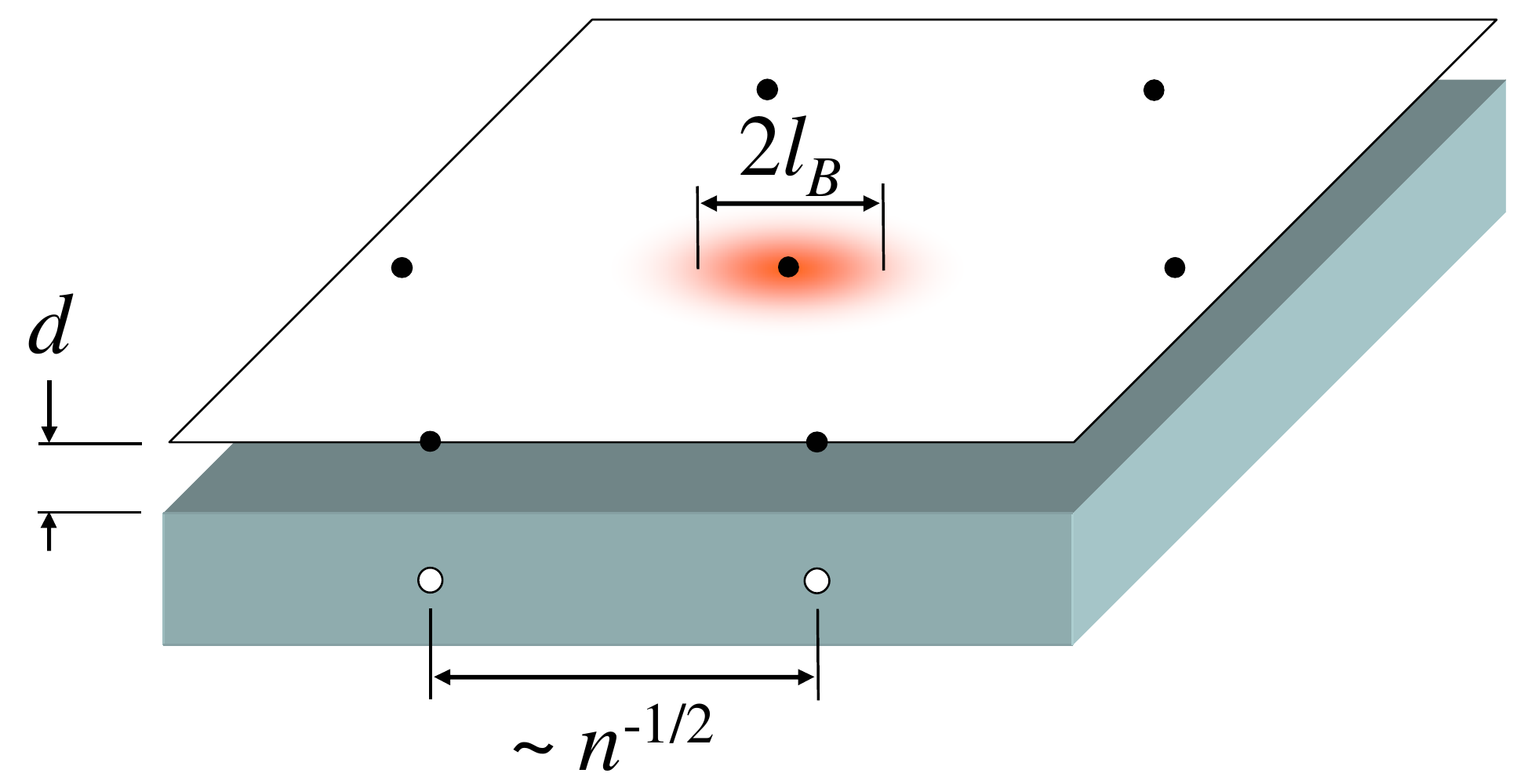}
\caption{(Color online) Schematic picture of a capacitor made with a low-density 2DEG next to a metal electrode in the presence of a transverse magnetic field.  Electrons (filled circles) create image charges (open circles) in the metal electrode.  The transverse magnetic field confines electrons to their lowest Landau level wavefunction with characteristic lateral size $\lb$, as shown schematically for the central electron by the red shaded area.} \label{fig:schematic}
\end{figure}

In a recent series of papers \cite{Skinner2010alc, Skinner2010svm}, we showed that when the electron density is sufficiently low that $nd^2 \ll 1$, the metal--2DEG capacitor can be thought of as a collection of electron-image charge dipoles (as shown schematically in Fig.\ \ref{fig:schematic}), whose interaction acquires the form $\sim e^2 d^2/r^3$ due to the strong screening effect of the metal.  As a result of this screening, the capacitance becomes greatly enhanced above the geometric value at $n d^2 \ll 1$ whenever the effective Bohr radius $a_B$ is much smaller
\footnote{There are reasons \cite{Skinner2010alc} to think that $a_B < d$ for the samples of Refs.\ \cite{Li2011vlc, Tinkl2012lne}.}
than $d$.  In the limit $n \ll a_B^2/d^4$, positional correlations between electrons are lost due to the short-ranged nature of the dipole interaction, and the capacitance saturates at a finite value such that $d^* = a_B/4$ (for spin-unpolarized electrons).  

One can expect, then, that for 2DEGs with large effective Bohr radius $a_B \gg d$, there can be no enhancement of the capacitance above the geometric value.  In particular, capacitors made from graphene, for which $a_B$ is effectively infinite due to the massless Dirac spectrum, seemingly should not produce $d^* < d$.  Such thinking is consistent with recent experiments probing the quantum capacitance of graphene \cite{Ponomarenko2010dos, Xia2009moq}, which showed that the capacitance is everywhere smaller than the geometric value and, in the absence of disorder, tends to zero in the limit $n \rightarrow 0$.  

In this paper, however, we show that even for 2DEGs with large (or infinite) Bohr radius, an enormous enhancement of the capacitance is possible when the capacitor is placed in a strong perpendicular magnetic field.  Such an applied field helps to confine electrons laterally and preserve their strong positional correlations even in the limit \cite{Lozovik1975ctd} $n \rightarrow 0$, so that the capacitance diverges.  This situation is illustrated schematically in Fig.\ \ref{fig:schematic}.  

\begin{figure}[htb!]
\centering
\includegraphics[width=0.5 \textwidth]{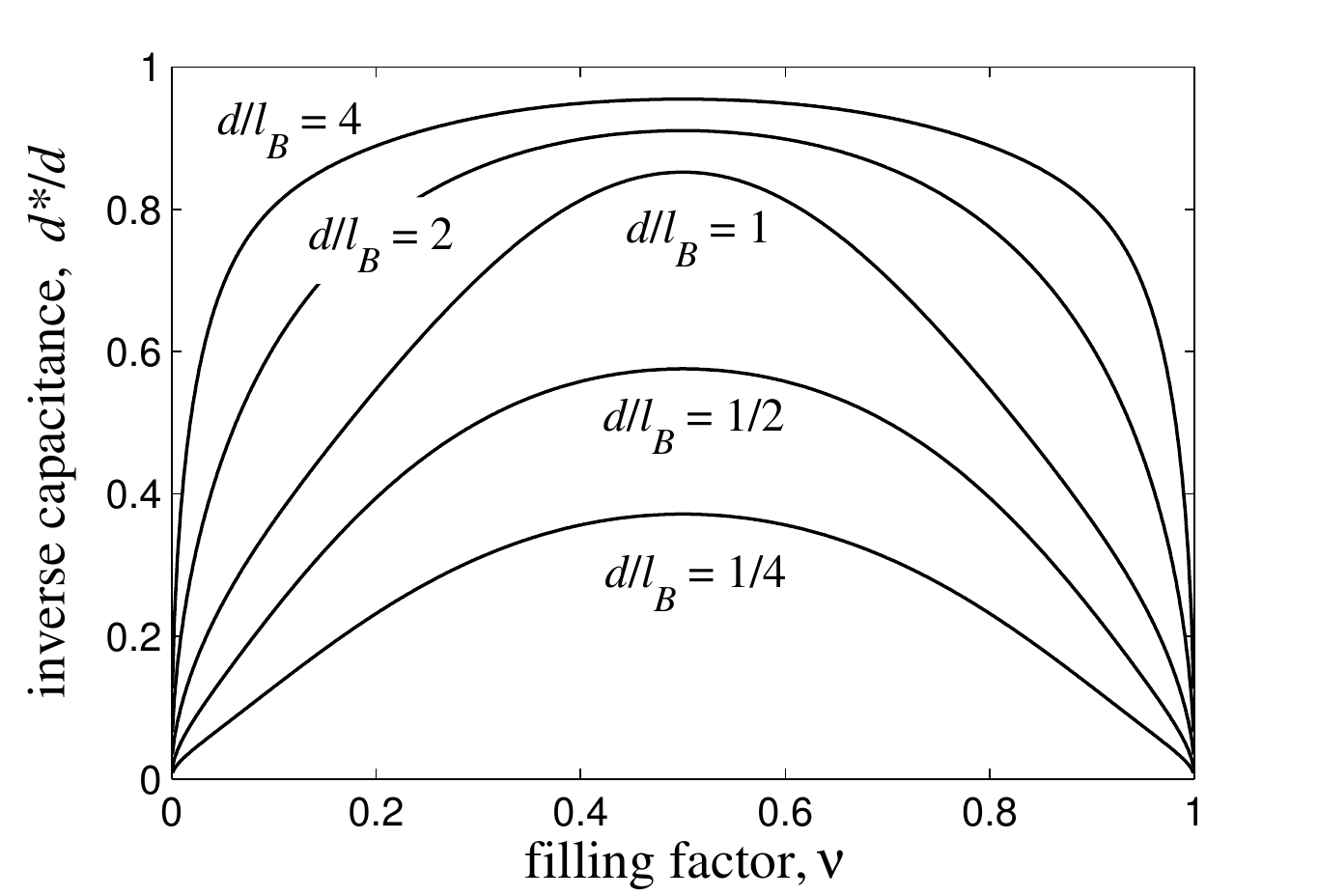}
\caption{The effective capacitor thickness $d^*$, in units of $d$, plotted as a function of filling factor $\nu$ in the LLL at different values of $d/\lb$.} \label{fig:dstar-intro}
\end{figure}

In this paper we focus our attention specifically on the lowest Landau level (LLL), where the filling factor $\nu < 1$, and we calculate $d^*$ as a function of $\nu$ in the absence of disorder.  Our primary result is illustrated in Fig.\ \ref{fig:dstar-intro}, which shows $d^*/d$ as a function of filling factor for different values of $d/\lb$, where $\lb = \sqrt{\hbar c/eB}$ is the magnetic length.  From Fig.\ \ref{fig:dstar-intro} one can see that when the capacitor is thin enough that $d/\lb \ll 1$, the capacitance is greatly enhanced above the geometric value throughout the LLL.  For large $d/\lb$, this enhancement is large only at $\nu \ll 1$ and $1 - \nu \ll 1$.  These results are valid for a capacitor made with either a conventional spin-polarized 2DEG or with a graphene layer where both spin and valley degeneracies are lifted, due to an exact correspondence between the two systems in the LLL.  We assume throughout this paper that we deal with such a nondegenerate LLL, so that the filling factor $\nu = 2\pi n \lb^2$.

In the remainder of this paper we first present our derivation of the capacitance (Sec.\ \ref{sec:general}), which is based on generalizing the Fano-Ortolani (FO) expression \cite{Fano1988ife} for the energy of a 2DEG in the LLL to account for screening by the gate electrode.  We then discuss, in Sec.\ \ref{sec:limits} the limit of large electron density, $nd^2 \gg 1$, at which the gate screening effect is unimportant, and we show that we can reproduce the traditional FO result.  The opposite limit $nd^2 \ll 1$ is also considered in detail, and we derive an analytical expression for $d^*$ that is valid throughout the LLL in cases where $d/\lb \ll 1$.  Throughout this paper we work in the range of temperatures where the thermal energy is small compared to the interaction energy per electron, yet still large enough to smear out the singularities in energy related to the fractional quantum Hall effect.  The effects of finite disorder are discussed in Sec.\ \ref{sec:disorder}.  Finally, in Sec.\ \ref{sec:two} we discuss capacitors made from two parallel, low-density 2DEGs, where very similar physics also produces large capacitance enhancement.

\section{General numerical result} \label{sec:general}

The general, thermodynamic definition of the differential capacitance is $C^{-1} = (1/e^2) d^2[n E(n)]/dn^2$, where $E(n)$ is the total energy per electron.  The effective thickness $d^*$ can thus be written
\be 
\frac{d^*}{d} = \frac{\varepsilon \lb^2}{2e^2d} \frac{d^2}{d\nu^2} \left[ \nu E(\nu) \right].
\label{eq:ddef} 
\ee
Particle-hole symmetry in the LLL \cite{Fano1988ife} implies that $E(\nu)$ must satisfy $\nu [E(\nu) - E(1)] = (1-\nu)[E(1-\nu) - E(1)]$.  As a consequence, $d^*$ in the LLL is symmetric about $\nu = 1/2$ (as can be seen in Fig.\ \ref{fig:dstar-intro}).  Given the particle-hole symmetry requirement, FO suggested that the energy in the LLL can be written using the following power law expansion:
\be 
\nu E(\nu) = E(1)\nu^2 + \sum_{k=0}^\infty \alpha_k [\nu (1-\nu)]^{k/2}.
\label{eq:EFO}
\ee
The condition $E(0) = 0$ implies that $\alpha_0 = \alpha_1 = \alpha_2 = 0$.

The energy $E(1)$ can be calculated \cite{Laughlin1990qhe} by integrating the electron-electron interaction law $V(r)$ weighted with the pair distribution function 
\be 
g_1(r) = 1 - \exp\left[-\frac{r^2}{2\lb^2} \right]
\label{eq:pdf}
\ee
of the incompressible state at $\nu = 1$.  For electrons with a nearby metallic gate, the interaction law is given by
\be 
V(r) = \frac{e^2}{\varepsilon r} - \frac{e^2}{\varepsilon \sqrt{r^2 + (2d)^2}},
\label{eq:Vr}
\ee
so that
\begin{eqnarray}
E(1) & = & \frac12 n \int d^2r V(r) g_1(r) 
\label{eq:E1} \\
& = & \frac{e^2}{\varepsilon \lb} \left[ \frac{d}{\lb} - \sqrt{\frac{\pi}{8}}\left(1 - \exp\left\{\frac{2d^2}{\lb^2} \right\} \text{erfc}\left\{\frac{\sqrt{2}d}{\lb} \right\} \right) \right]. \nonumber
\end{eqnarray}
In the first equality of Eq.\ (\ref{eq:E1}) we have substituted $n = 1/2\pi\lb^2$.

Thus, in order to produce an approximate analytical expression for the energy per electron $E(\nu)$ in the LLL it is sufficient to derive values for the coefficients $\alpha_k$ in Eq.\ (\ref{eq:EFO}).  The effective thickness $d^*$ can then be calculated using Eq.\ (\ref{eq:ddef}).  Of course, this approach cannot capture the small cusps in the energy associated with fractional quantum Hall states, which at finite temperature produce weak local maxima in $d^*$ \cite{Fano1988ife, Eisenstein1992nco, Eisenstein1994cot}.  Nonetheless, it does accurately reproduce the energy of the fractional quantum Hall states themselves, as we show below.

Our approach to estimating the coefficients $\alpha_k$ is as follows.  For a given value of $d/\lb$ and $\nu \ll 1$ we calculate using a numeric sum the Coulomb energy associated with a classical, triangular Wigner crystal of electrons interacting with the interaction law $V(r)$.  We also calculate the leading-order correction to this classical energy by using first-order perturbation theory for each electron in the slowly-varying Coulomb potential created by its neighbors.  This potential is expanded to lowest order in the distance $r$ from the potential energy minimum, and we use the LLL wavefunction $\psi(r) = \exp[-r^2/4\lb^2] / \sqrt{2\pi\lb^2}$.  The resulting estimate for $E(\nu)$ is calculated for a particular range of small filling factors, $0 < \nu < \nuc$, with $\nuc \ll 1$, at which positional correlations are strong and the energy of the Wigner crystal state is a reasonable approximation for the total energy.  Our choice of $\nuc$ is discussed more fully below.  Once $E(\nu)$ is known in the range $0 < \nu < \nuc$, the function $E(\nu)$ is then fit to the form of Eq.\ (\ref{eq:EFO}) using a polynomial best fit with $\alpha_{k < 3}$ and $\alpha_{k > 7}$ set to zero.  The energy $E(1)$ is given analytically by Eq.\ (\ref{eq:E1}).  From the coefficients $\alpha_k$ the inverse capacitance $d^*/d$ is evaluated using Eq.\ (\ref{eq:ddef}).  The results of this procedure are shown as the black solid lines in Figs.\ \ref{fig:dstar-intro}.

One reasonable estimate for the value of $\nuc$ is $\nuc = 0.2$, since $0 < \nu < 0.2$ corresponds roughly to the Wigner crystal regime in the unscreened 2DEG \cite{Lam1984lta}.  While the screened 2DEG considered here may not remain in a truly solid phase up until $\nu = 0.2$, one can nonetheless expect that local positional correlations persist strongly, so that the WC state remains an accurate approximation to the total energy per electron.  Fig. \ref{fig:dstar-intro} uses $\nuc = 0.2$, but we find that our numerical results for $d^*$ are not noticeably different even if $\nuc$ is made as small as $0.05$.

\begin{figure}[htb!]
\centering
\includegraphics[width=0.5 \textwidth]{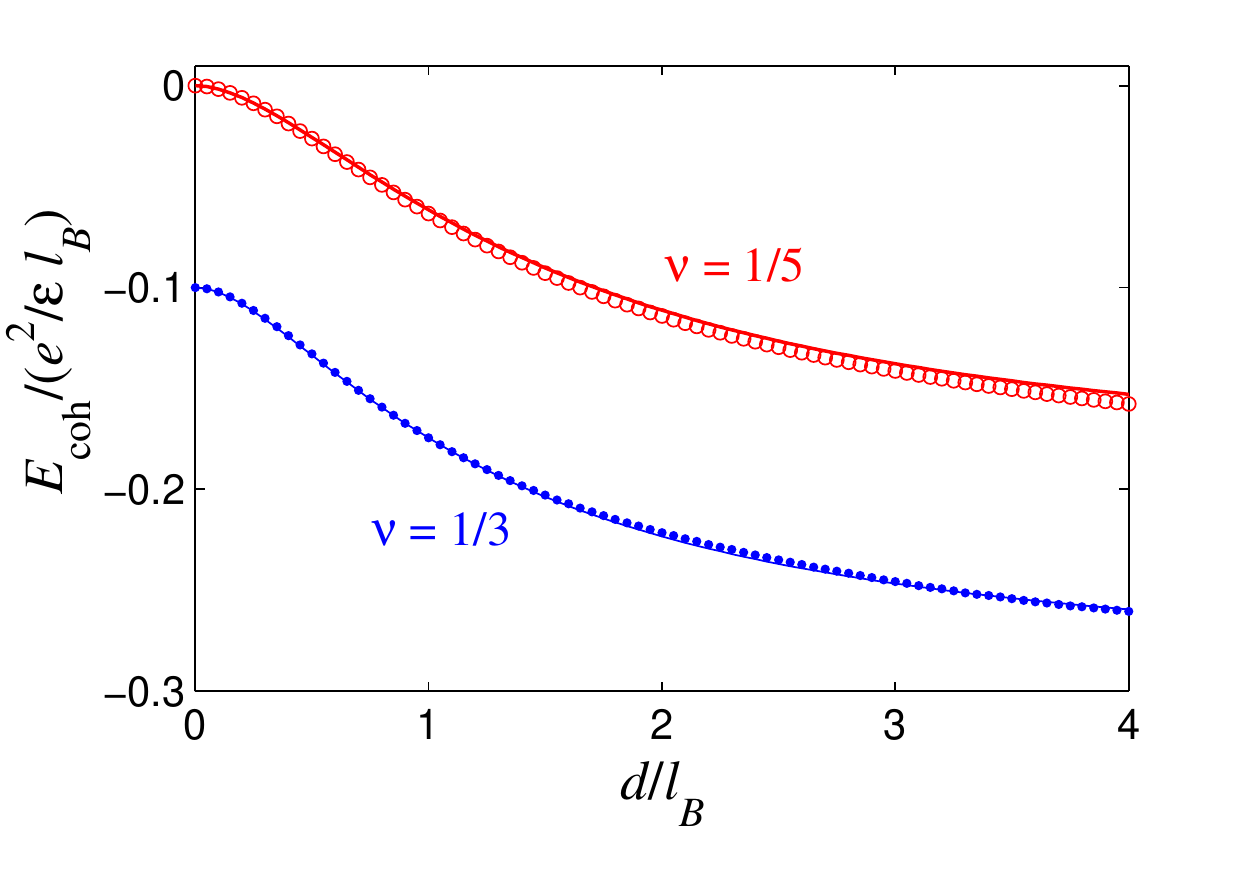}
\caption{(Color online)  The cohesive energy per electron at filling factors $\nu = 1/3$ (lower curve, closed symbols) and $\nu = 1/5$ (upper curve, open symbols) as a function of $d/\lb$.  Symbols show the results of the interpolation procedure described in Sec.\ \ref{sec:general} and the curves show the result of Eq.\ (\ref{eq:Eg}) using the parameterizations of $g_{1/3}(r)$ and $g_{1/5}(r)$ in Ref.\ \onlinecite{Girvin1986mrt}.  For clarity, the curve and symbols corresponding to $\nu = 1/3$ have been displaced downward by $0.1$.} \label{fig:Ecoh}
\end{figure}

We also checked that our interpolation technique is capable of accurately reproducing the energy of the fractional quantum Hall states at $\nu = 1/3$ and $\nu = 1/5$.  For these states the pair distribution functions [$g_{1/3}(r)$ and $g_{1/5}(r)$, respectively] have been parameterized from Monte Carlo data \cite{Girvin1986mrt}, and one can calculate the energy 
\be 
E(\nu) = \frac{\nu}{4 \pi \lb^2} \int d^2r V(r) g_{\nu}(r)
\label{eq:Eg}
\ee 
directly without resorting to interpolation \cite{Fogler1997llt}.  Comparing the result of this calculation to the result we obtain from our interpolation procedure shows a close agreement at all values of $d/\lb$ for both $\nu = 1/3$ and $\nu = 1/5$.  This is demonstrated in Fig.\ \ref{fig:Ecoh}, where we plot the cohesive energy per electron, $\ecoh(\nu) = E(\nu) - E(1)\nu$, as a function of $d/\lb$.  In both cases our interpolation yields a result for $\ecoh$ that is everywhere identical to that of Eq.\ (\ref{eq:Eg}) to within a few percent.

\section{Analytical theory for two limiting cases} \label{sec:limits}

The general procedure described above allows one to numerically calculate $d^*$ within the LLL at arbitrary values of $\nu$ and $d/\lb$.  We now show that in the asymptotic limits $nd^2 \gg 1$ and $nd^2 \ll 1$, $d^*(\nu)$ can readily be described analytically.  For the case $nd^2 \gg 1$, where the average distance between electrons is much smaller than $d$, electron-electron interactions are not significantly screened by the metal electrode.  In this limit one can think that the system consists of a normal, unscreened 2DEG with a uniform neutralizing background that is displaced by a distance $d$.  By Eq.\ (\ref{eq:Cdef}), the corresponding effective thickness can be written 
\begin{eqnarray} 
\frac{d^*}{d} & = & 1 + \frac{\varepsilon \lb^2}{2e^2d} \frac{d^2}{d\nu^2} \left[ \nu \efo(\nu) \right]  
\label{eq:dFO} \\
& & \text{when } d/\lb \gg 1 \text{ and } \nu, 1-\nu \gg \lb^2/d^2, \nonumber
\end{eqnarray}
where $\efo$ is the FO expression \cite{Fano1988ife} for the energy per electron of the unscreened 2DEG with a coplanar neutralizing background.  
A form of Eq.\ (\ref{eq:dFO}) was first proposed by Efros \cite{Efros1990dos} and used to explain the observed negative compressibility of 2DEGs in semiconductor heterostructures \cite{Eisenstein1992nco, Eisenstein1994cot}.  As we show below, however, Eq.\ (\ref{eq:dFO}) becomes incorrect if pushed to the limit where corrections to the geometric capacitance are large.

\begin{figure}[htb!]
\centering
\includegraphics[width=0.5 \textwidth]{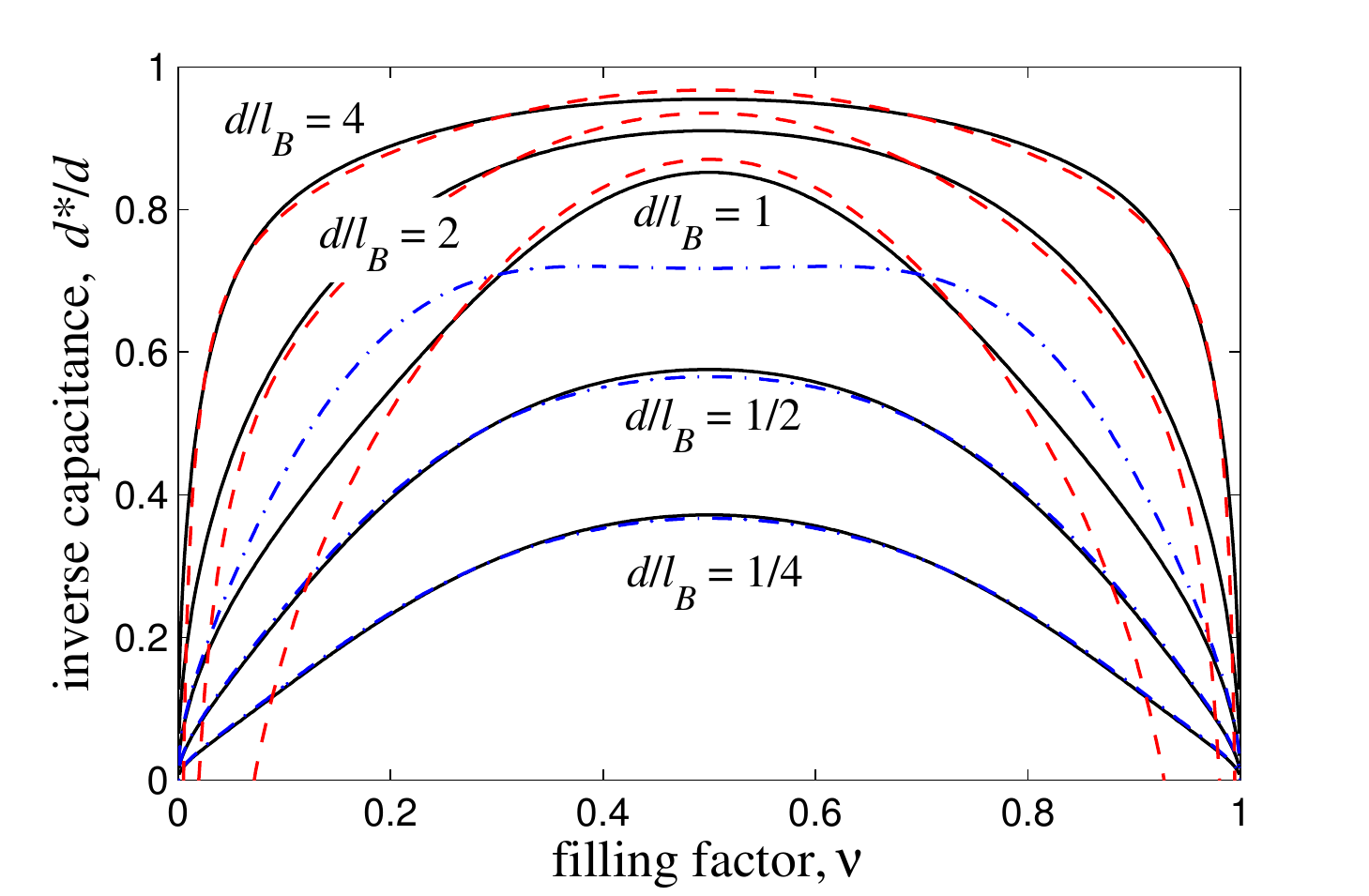}
\caption{(Color online) The effective capacitor thickness, in units of $d$, as a function of filling factor in the LLL for various values of $d/\lb$.  The result of Eq.\ (\ref{eq:dFO}) is plotted as the red dashed lines, and the blue dash-dotted lines correspond to the coefficients given by Eq.\ (\ref{eq:alpha}).  The solid black lines are the result of the general numerical fitting procedure described in Sec.\ \ref{sec:general}.} \label{fig:dstar}
\end{figure}

We note that if one applies our method outlined above for calculating the coefficients $\alpha_k$ and the energy $E(1)$ to the case of the unscreened 2DEG, one arrives at an expression for $E(\nu)$ that is everywhere equal to $\efo(\nu)$ to within $3.5 \%$.  Our semiclassical approach for estimating $E(\nu)$ at small $\nu$ also reproduces the Hartree-Fock expression \cite{Lam1984lta, Yoshioka1983geo, Maki1983sad} up to a term of order $\nu^{5/2}$.  
Eq.\ (\ref{eq:dFO}) is plotted for a few different values of $d/\lb$ as the red dashed lines in Fig.\ \ref{fig:dstar}, and indeed it matches our result in the appropriate limits $d/\lb \gg 1$ and $\nu, 1-\nu \gg (\lb/d)^2$.  

It should be noted that if Eq.\ (\ref{eq:dFO}) is naively pushed to the limit $\nu \ll (\lb/d)^2$, it yields a negative capacitance.  Such negative capacitance is forbidden by thermodynamic stability arguments \cite{Landau1984eoc}, and this failure of the FO expression comes because at very small $\nu$ the interaction between electrons is screened by the metal gate \cite{Skinner2010alc}.

Instead, in the limit $nd^2 \ll 1$ the ground state can be described as a triangular Wigner crystal of electron-image charge dipoles, as shown schematically in Fig.\ \ref{fig:schematic}, with interaction law $V(r) \simeq 2 e^2 d^2/\varepsilon r^3$.  Treating these electrons classically and summing their interactions, as described above, gives an energy per electron $E(\nu) \simeq 0.564 (e^2 d^2/\lb^3) \nu^{3/2}$.  Adding the leading-order correction to this expression from first-order perturbation theory gives 
\begin{eqnarray} 
E(\nu) & \simeq & \frac{e^2 d^2}{\lb^3} (0.564 \nu^{3/2} + 0.429 \nu^{5/2}) 
\label{eq:Enu0} \\
& & \text{when } \nu \ll \min \left\{1, \lb^2/d^2 \right\}.
\nonumber
\end{eqnarray}
This expression can be used to determine the coefficients $\alpha_k$ for $3 \leq k \leq 7$ by matching orders of $\nu$ to Eq.\ (\ref{eq:EFO}), which gives:
\begin{eqnarray}
\alpha_3 & = & 0 \nonumber \\
\alpha_4 & = & - E(1) \nonumber \\
\alpha_5 & = & 0.564 e^2 d^2/\lb^3 
\label{eq:alpha} \\
\alpha_6 & = & - 2E(1) \nonumber \\
\alpha_7 & = & 1.84 e^2 d^2/\lb^3 \nonumber \\
& & \text{when } d/\lb \ll 1 \text{ or } \nu, 1-\nu \ll \lb^2/d^2 \nonumber .
\end{eqnarray}

Equations (\ref{eq:ddef}), (\ref{eq:EFO}), and (\ref{eq:alpha}) can be combined to produce an analytical expression for $d^*$.  This is plotted as the blue dash-dotted lines in Fig.\ \ref{fig:dstar}.  (Due to its cumbersome algebraic form, we do not write this expression explicitly here.)  For cases where $d/\lb \lesssim 1/2$, $d^*$ is well described by Eqs.\ (\ref{eq:ddef}), (\ref{eq:EFO}), and (\ref{eq:alpha}) for the entire range $0 < \nu < 1$.  For larger $d/\lb$, this description is accurate only when $\nu$ is very close to zero or unity.
In the limit $\nu \rightarrow 0$, one gets $d^*/d \simeq 1.06 (d/\lb) \sqrt{\nu}$, which is equivalent to the classical result derived in Ref.\ \onlinecite{Skinner2010alc}.  Unlike in Ref.\ \onlinecite{Skinner2010alc}, however, this result is correct even for 2DEGs with large Bohr radius, due to the confining effects of the transverse magnetic field.

Overall, our primary result shown in Figs.\ \ref{fig:dstar-intro} and \ref{fig:dstar} is striking: it suggests that in the presence of a transverse magnetic field the capacitance can be arbitrarily large, limited only by temperature or the presence of a disorder potential.  This is particularly surprising for graphene, where in the absence of magnetic field the capacitance becomes vanishingly small \cite{John2004qci, Ponomarenko2010dos, Xia2009moq} at $\nu \rightarrow 0$.  This enormous enhancement of the capacitance in a magnetic field is a sister effect to Wigner crystallization in graphene, which is not possible in the absence of a magnetic field due to graphene's linear spectrum \cite{Dahal2006aow}, but which in the presence of a field becomes possible \cite{Zhang2007wca} at small $\nu$.

\section{Disorder effects} \label{sec:disorder}

\begin{figure}[htb!]
\centering
\includegraphics[width=0.45 \textwidth]{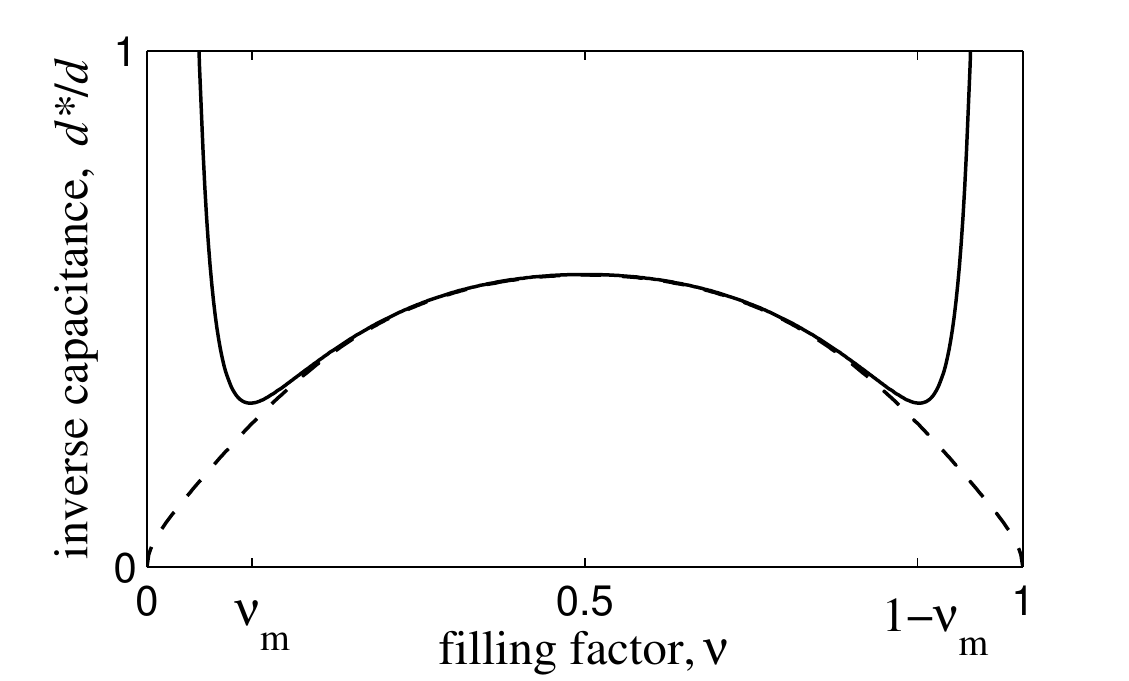}
\caption{Schematic illustration of the effect of disorder on the measured inverse capacitance.  The dashed line shows $d^*/d$ in the absence of disorder, and the solid line shows $d^*/d$ in the presence of a finite but small disorder potential, for which $\num \ll 1$.} \label{fig:disorder-schematic}
\end{figure}

Of course, our analysis so far has ignored the effects of disorder, which truncate the divergence of the capacitance by destroying positional correlations.  The presence of disorder modulates the density of the correlated electron liquid with some characteristic amplitude $\delta n_d$.  At small enough average electron density that $n < \delta n_d$, pores open up in the 2DEG and electric field lines starting at the metal gate leak through these pores.  As a result, the electronic compressibility rapidly becomes small and $d^*$ grows sharply at small $\nu$ and small $1 - \nu$, as observed in Refs.\ \onlinecite{Eisenstein1992nco, Eisenstein1994cot} and illustrated schematically in Fig.\ \ref{fig:disorder-schematic}.  Thus, observation of the large capacitance enhancement predicted here requires devices that are thin enough and clean enough that the characteristic disorder impurity concentration $\delta n_d$ satisfies $\delta n_d \ll 1/d^2$.  For the very thin devices examined in Ref.\ \onlinecite{Gorbachev2012scd}, where $d \sim 5$ nm, this implies $\delta n_d \ll 4 \times 10^{12}$ cm$^2$.  

For devices with such small disorder, our predictions for $d^*$ should be correct up to some particular small value of $\nu = \num$, at which $n$ becomes similar in magnitude to $\delta n_d$, and $d^*$ rises sharply.  This behavior is illustrated schematically in Fig.\ \ref{fig:disorder-schematic} (which resembles figure 2 of Ref.\ \onlinecite{Eisenstein1992nco}).  A previous work \cite{Fogler2004nsa} has shown that the minimum in the electronic compressibility for the unscreened 2DEG occurs at $n \approx 0.4 \delta n_d$, which implies
\be 
\num \approx 2.4 (\delta n_d) \lb^2.
\ee
Thus, in the window of filling factors $\num < \nu < 1-\num$, the effective thickness is small and well-described by the theories presented here, while at smaller or larger $\nu$ within the LLL $d^*$ is large.  If the disorder is large enough (or the magnetic field is small enough) that $(\delta n_d)/B \gtrsim 3 \times 10^{10}$ cm$^{-2}$/T, this window closes completely, and the capacitance in the LLL is everywhere determined by the disorder.

\section{Capacitor of two  2D gases} \label{sec:two}

So far we have focused on the case of a single 2DEG adjacent to a metal electrode. In this section we discuss capacitors made from two parallel 2DEGs, and show that similar capacitance enhancement effects can appear in a strong magnetic field.  As a trivial example, one can notice that if one of the 2DEGs is sparse while the other has a much larger electron concentration, then this second 2DEG is equivalent to a metal electrode in terms of its electronic screening properties, and the theory outlined above is applicable.  On the other hand, one can consider a capacitor made from a sparse 2DEG and a parallel, 2D \emph{hole} gas with equal hole concentration $n$ (as in Refs.\ \onlinecite{Sanchez-Yamagishi2012qhe, Gorbachev2012scd, Ponomarenko2011tmt}).  In such devices, for sufficiently small $d$ (and in the LLL), electrons in one layer bind strongly to holes in the opposite layer, forming indirect excitons that interact as dipoles \cite{Lozovik1975fsp}.

The electronic state of such a system was described by Yoshioka and MacDonald \cite{Yoshioka1990dqw}, who noticed that a sparse \emph{electron-hole} bilayer system with total filling factor zero is equivalent to a bilayer \emph{electron} system with total filling factor one.  That is, a system in which one layer contains electrons occupying a partially-filled LLL, $\nu < 1$, and the other layer contains an equal number of holes is equivalent to a system in which one layer has filling factor $\nu < 1$ of electrons and the other layer has filling factor $1 - \nu < 1$ of electrons.  These authors showed that when the distance between the two layers is small enough that $d/\lb \ll 1$, the electrons in the two layers together effectively occupy the incompressible ground state wavefunction \cite{Halperin1983tqh} $\Psi_{1,1,1}$, even though the constituent electrons are spread between two layers.  This result has since been confirmed by more detailed numerical studies \cite{Chen1991ecd, Zhang2007sie}.  The corresponding pair distribution function between electrons is therefore identical to that of the $\nu = 1$ incompressible state, Eq.\ (\ref{eq:pdf}), so that the interaction ($n$-dependent) part of the total energy per electron can be calculated as $E_\text{int}(\nu) = \frac12 n \int d^2r V(r) g_1(r)$, as in Eq.\ (\ref{eq:E1}), with $V(r) = 2(e^2/\varepsilon r - e^2/\varepsilon \sqrt{r^2 + d^2})$ and $n = \nu/2 \pi \lb^2$.  Taking derivatives of this result for $E_\text{int}$\cite{Yoshioka1990dqw, Chen1991ecd, Yang2001des, Joglekar2002bvi} gives, according to Eq.\ (\ref{eq:ddef}),
\be 
\frac{d^*}{d}  = 1 - \sqrt{\frac{\pi}{2}} \frac{\lb}{d} \left(1 - \exp \left [ \frac{d^2}{2 \lb^2} \right] \text{erfc} \left[ \frac{d}{\sqrt{2} \lb} \right] \right).
\label{eq:dsdoublelong}
\ee
In the limit $d/\lb \ll 1$, this expression becomes
\be 
\frac{d^*}{d} \simeq \sqrt{\frac{\pi}{8}} \frac{d}{\lb}
\label{eq:dsdouble}.
\ee 

One can notice that this result for $d^*/d$ is similar in magnitude to the results presented in Sec.\ \ref{sec:limits} for the middle of the LLL; in both cases $d^*/d \propto d/\lb$ at small $d/\lb$.  
In other words, capacitor devices comprised of two electron or hole gases also exhibit capacitance that is greatly enhanced over the geometric value when $d/\lb \ll 1$ and the LLL is partially filled.  As mentioned above, this result is unchanged if one considers electrons/holes in graphene rather than in a conventional semiconductor quantum well, where electrons have a quadratic dispersion relation, due to the exact equivalence between the wavefunctions of these two systems in the LLL.

Unlike the results shown in Fig.\ \ref{fig:dstar-intro}, Eq.\ (\ref{eq:dsdouble}) describes a constant, $\nu$-independent capacitance in the LLL.  On the other hand, when $d/\lb$ is of order unity, collective excitations about the $\Psi_{1,1,1}$ state (electron correlations) lead to ``softening" of the incompressible state and produces $\nu$-dependent corrections \cite{Joglekar2002bvi} to the result of Eq.\ (\ref{eq:dsdoublelong}).  These corrections become larger with increasing $d/\lb$, and at $d/\lb \gg 1$ electrons and holes in opposite layers decouple from each other \cite{Chen1991ecd, Joglekar2002bvi, Zhang2007wca}.  A careful calculation of the capacitance for double-2DEG capacitors at such $d/\lb \gtrsim 1$ is left as the subject for a later work.

Finally, we note that large capacitance enhancement may also be seen in situations where two parallel graphene layers with separation $d$ are electrically connected (have the same electrochemical potential) and are placed in a perpendicular displacement field $\vec{D}$ created by two distant metallic gates that are parallel to the graphenes \cite{Sanchez-Yamagishi2012qhe, Gorbachev2012scd}. In this case, the system of two layers is polarized along $\vec{D}$.  In a strong magnetic field this polarization takes the form of discrete, parallel, mutually repulsive dipoles, the charges of which are confined within disks of  radius $\lb$ within the graphene planes.  Using our theory above one can find the 2D concentration of dipoles $n$ as a function of the ``applied voltage" $V=|\vec{D}|d$, as well as the effective ``capacitance per unit area" $C =  e dn/dV$, or, equivalently, the corresponding ratio $d^*/d$.  This ratio is given by exactly the same expression as for a real capacitor made from two layers of graphene and charged by a real voltage (electrochemical potential difference), as we described above. In the case of electrically connected graphene layers it is possible to deal with very small\cite{Sanchez-Yamagishi2012qhe, Gorbachev2012scd} $d$ and, therefore, potentially arrive at a giant capacitance enhancement.  Substantial enhancement was already seen in Ref.\ \onlinecite{Sanchez-Yamagishi2012qhe}.

\acknowledgments

We are grateful to A. L. Efros, M. M. Fogler, Y. Joglekar, A. H. MacDonald, and K. S. Novoselov for useful discussions.
This work was supported primarily by the National Science Foundation through the University of Minnesota MRSEC under Award Number DMR-0819885.

\bibliography{MC}

\end{document}